\documentclass[conference]{IEEEtran}
\IEEEoverridecommandlockouts
\usepackage{cite}
\usepackage{amsmath,amssymb,amsfonts}
\usepackage{algorithmic}
\usepackage{graphicx}
\usepackage{textcomp}
\usepackage{float}
\usepackage{caption}
\usepackage{subcaption}
\usepackage{balance}

\usepackage{xcolor}
\def\BibTeX{{\rm B\kern-.05em{\sc i\kern-.025em b}\kern-.08em
    T\kern-.1667em\lower.7ex\hbox{E}\kern-.125emX}}

\begin{document}

\title{A Survey on Machine Learning Algorithms for Applications in Cognitive Radio Networks}
\author{\IEEEauthorblockN{Akshay~Upadhye,  Purushothaman~Saravanan, Shreeram~Suresh~Chandra and Sanjeev~Gurugopinath}
\IEEEauthorblockA{Department of Electronics and Communication Engineering, PES University, Bengaluru 560085, India.\\
Emails: \{upadhyeakshay2, purushothaman.syadav, shreeramsureshchandra\}@gmail.com, sanjeevg@pes.edu}}

\maketitle

\begin{abstract}
In this paper, we present a survey on the utility of machine learning (ML) algorithms for applications in cognitive radio networks (CRN). We start with a high-level overview of some of the major challenges in CRNs, and mention the ML architectures and algorithms that can be used to alleviate them. In particular, our focus is on two fundamental applications in CRNs, namely spectrum sensing -- with non-cooperative and cooperative scenarios, and dynamic spectrum access -- with spectrum auction and prediction. We present a detailed study of recent advancements in the field of ML in CRNs for these applications, and briefly discuss the set of challenges in real-time implementation of ML techniques for CRNs.
\end{abstract}


\section{Introduction}\label{SecIntro}
As the number of devices connected to the internet increases day by day, there is a severe constraint on the allocation of spectrum resources. It is a well-known fact that spectrum is a limited resource. Spectrum bands are open auctioned, and are sold to the highest bidder. Broadly speaking, communication spectra are  divided into licensed and unlicensed bands. The licensed user has access to the channel at all times. However, when the licensed user does not use it, the channel is available for communication. According to the federal communications commission (FCC), spectrum under utilization is one of the main causes of this spectrum scarcity. This gave rise to a new generation of edge devices called cognitive radios (CR). A CR is an enhanced software-defined radio, which learns from its environment, and can also adjust to changes in the nature of licensed signals, network parameters, and Qos requirements. Primary applications of CR include communication systems that require high reliability whenever and wherever needed \cite{1391031}. CRs extend the functions of a software radio. %with radio-domain model-based reasoning about a network etiquette.
A major hurdle in a wireless network is efficient spectrum management. Even with significant investment in 3G infrastructure, the spectrum allocated to 3G was limited. To overcome this, CR offers a new class of protocols called \emph{formal radio etiquettes} \cite{819467} for the flexible pooling of radio spectrum resources. Similarly, the works in \cite{6507396,7945403,5304033} have explored the design of CR in a 4G network. In recent times, with the advent of 5G and its close interlude with internet-of-things (IoT), the number of devices connecting to the internet is now more than ever. Hence the need to inculcate CR networks to utilize the spectrum effectively has become an important research topic. Although several classical signal processing methods exist for CR applications like spectrum sensing, dynamic spectrum access, decision-making, and spectrum prediction, they are model-based and their performance degrades with uncertainties in model \cite{8869190}. On the other hand, ML techniques exploit the experimentally captured data, and when trained properly across a variety of data, can robustly learn the radio environments, adapt to new conditions and consistently yield a good performance. These algorithms are modular in their architectures and hence form elegant solutions to problems encountered in CR networks \cite{6336689}.

In this paper, we present a contemporary survey on the utility of ML algorithms for CR applications. As mentioned earlier, ML algorithms are envisioned to solve several problems in CRNs, thanks to the rapid development in ML techniques and architectures. First, we present a high-level overview of some of the key challenges in CRNs and mention the architectures and algorithms in ML that can be used to alleviate them. Next, given the vast scope of approaches for CR, we focus our attention on two fundamental applications, namely spectrum sensing (SS), dynamic spectrum access (DSA) and subtopics such as cooperative sensing, spectrum prediction and spectrum auction. We provide a detailed description of recent advancements in the field of ML for CR for the above mentioned applications. Moreover, we briefly discuss a set of challenges in real-time implementation of the mentioned ML techniques.

The remainder of this paper is organized as follows. Some of the key challenges in CRNs and corresponding ML-based solutions are presented in Section~\ref{SecKeyChallenges}. Section~\ref{SecSS} details the SS problem in CRNs, and presents the popular ML algorithms for conventional and cooperative SS. In Section \ref{SecDSA}, DSA is studied in detail as a learning problem, and existing ML solutions for problems such as spectrum auction and spectrum prediction are discussed in detail. Common challenges encountered while employing ML techniques for CR are briefed in  Section~\ref{SecChall}, and concluding remarks are provided in Section~\ref{conc}.

\begin{figure*}[t]
    \centering
    \includegraphics[height=6cm, width=15cm]{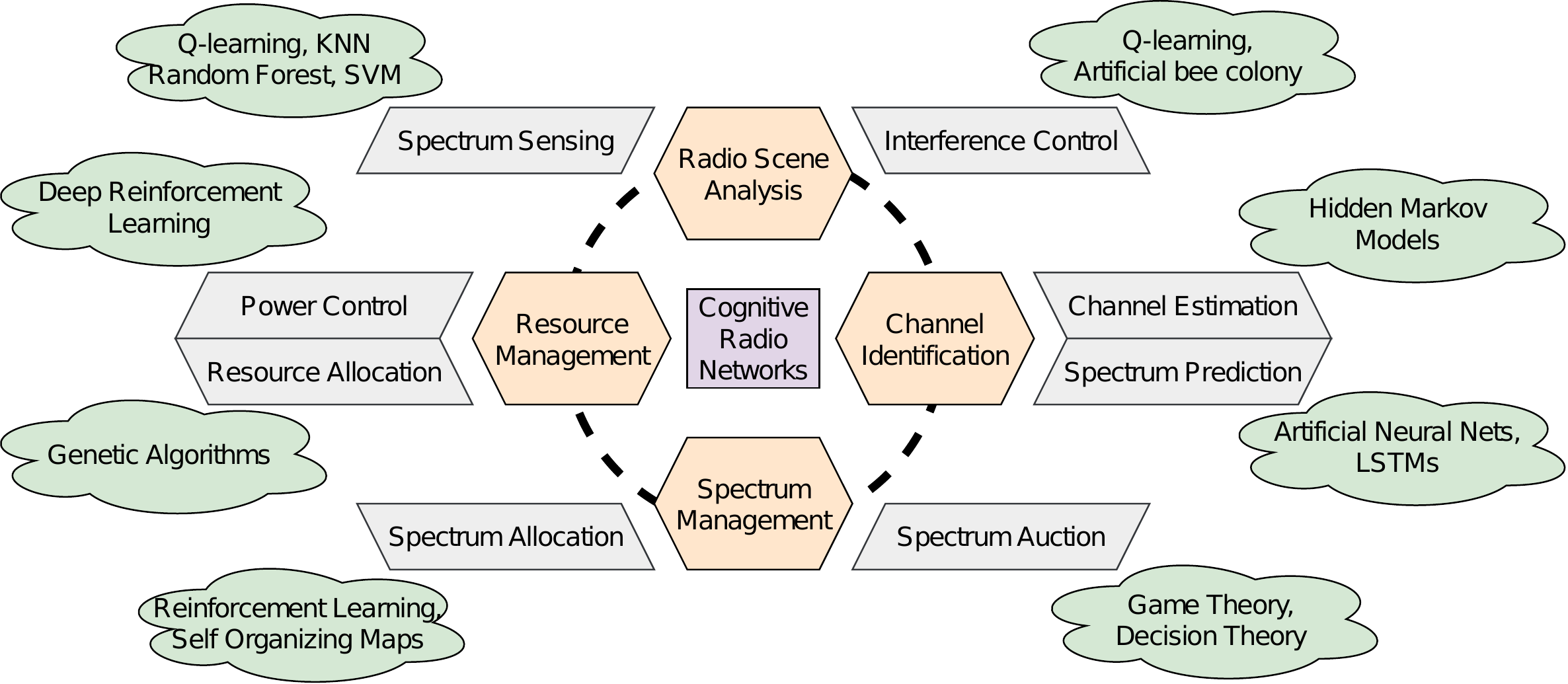}
    \caption{Key challenges in cognitive radios and the corresponding machine learning-based solutions.}
    \label{fig:mlincr}
    \hrulefill
\end{figure*}

\section{Key Challenges and ML Solutions} \label{SecKeyChallenges}
In this section, we discuss some of the major challenges in designing a CR network, and mention some of the ML-based solutions to alleviate them. At a high level, these key challenges in CRNs can be categorized into four classes, namely, radio scene analysis, channel identification, spectrum management and resource management, as shown in Fig.~\ref{fig:mlincr}. A CR needs to be aware of the radio frequency activity in its environment. In radio scene analysis, SS and interference control are the two important tasks. The key problem in SS is to detect white spaces, while in interference control, the requirement is to avoid harmful interference with a primary user (PU). Distributed Q-learning has been proposed to control the interference caused by multiple CRs on a PU \cite{5415567}. Once these white spaces are detected, DSA is necessary to optimize the spectrum usage. Game theory and reinforcement learning are some of the popular techniques used in spectrum management \cite{6554535, 5692880}. In resource management, the goal is to manage the resources in an efficient way to meet the quality-of-service (QoS) requirements. Genetic algorithms and reinforcement learning are some of the ML algorithms to tackle this problem \cite{7173287}. Frequently used algorithms in channel estimation and spectrum prediction are hidden Markov model and long short term memory (LSTM), respectively. Other ML algorithms for various tasks under these four categories are mentioned in Fig.~\ref{fig:mlincr}. Due to the vast advancements in the field of ML for CR, we restrict our attention to ML techniques for SS and DSA in this paper. In the next section, we discuss the advancements in SS in detail.

\section{Spectrum Sensing} \label{SecSS}
Spectrum sensing is one of the most essential functionalities in a CR, which is the task of deciding the presence or absence of a PU. Many classical signal processing methods such as energy detector (ED), matched filter detector (MFD), cyclostationary feature-based detector, and differential entropy detector (DED) have been proposed \cite{5604136, 7466780, 5670882}. However, there are some limitations to these methods. In the case of ED, the variance of the noise has to be known a priori in order to set the threshold. The MFD assumes that the CR has complete information about the primary signal, while the cyclostationary detector is computationally complex. To overcome these problems, several ML algorithms have been proposed \cite{8830434, 7986442, 8834099, 7852915, 9020304, 8830453, 8292449}. In an ML framework, SS can be framed as a classification problem \cite{8850035}. The received signal power and the correlation of the cyclic prefix were used as inputs to train a naive Bayes' classifier in \cite{8850035}. Cyclic prefix induced correlation was used to overcome the signal-to-noise-ratio (SNR) wall phenomenon in ED in \cite{5404390, 7756405}. 

Deep convolutional neural networks (CNN) for SS were proposed in \cite{9020304, 9187593, 8685112, 8830453}. Similarly, a deep learning classification technique was proposed in \cite{9020304}. 
The power spectrum of the received signal was used as the input to train the CNN model. Normalization of the input signal power was performed to solve the problem of noise uncertainty. In order to generalize the model, it was trained on different signals. Different types of modulation schemes like BPSK, QPSK, FSK, QAM, and PAM were used for training. A four-layer CNN was developed in \cite{9187593} to restrict the computational cost while meeting the SS requirements. The proposed method was evaluated against a dataset collected in the ultrahigh frequency TV band. The aim of the proposed method was to decrease the training duration in cases where the input signal features change. Transfer learning was used to decrease the training duration, as it enables a swift adaption to changes in frequency, environment, and location. When the training time is reduced, sensing time is also reduced which directly improves the overall system throughput. The spectrum of the received signal was used for training the CNN to obtain the features. Features learned by the deep CNN were passed as inputs to a support vector machine with a linear kernel to classify the spectrum. The probability of detection and probability of false-alarm were computed and evaluated to confirm that they fulfill the sensing requirement. Performance analysis of the suggested method was performed with and without transfer learning. It was shown that there was a decrease in spectrum sensing time when transfer learning was employed.

Exploiting the knowledge of history of the activity of a PU, \cite{8685112} uses a CNN-based deep learning algorithm for activity pattern aware spectrum sensing (APASS). This approach was data-driven and did not assume any information about signal statistics and PU activity pattern model. For model-based frameworks, the sensing performance depends on the accuracy of the considered models. However, when used on real incoming data, there can be mismatches and the performance of the above methods can decline significantly. No such assumptions are made in data-driven frameworks. These methods learn from the pattern present in the input. This helps in reducing the effect of model-mismatch problems. Two phases are proposed in data-driven detectors, namely (i) offline training phase, and (ii) an on-line detection phase. In \cite{8685112}, the correlation matrix (CM) was used as a test statistic for SS, since it contains information of the energy of the signal and correlation among the signals. Interestingly, this can also be interpreted as an image and passed to CNN as inputs for training. To capture the activity of a PU, the \emph{present} CM and \emph{past} CM of the PU signal were considered for training. These inputs were passed through several convolution and max-pooling layers. The output after these layers were flattened and concatenated. After applying a softmax regression, sensing results were interpreted. The APASS algorithm has the advantage that it can learn the PU activity pattern with no prior assumption on the PU signal and the signal model. Next, we look at some of the methods proposed for cooperative spectrum sensing (CSS).

\subsection{Cooperative Spectrum Sensing} \label{SecCSS}
In conventional SS, a single CR senses the spectrum for channel availability. Two main problems occur because of this approach: (a) the hidden node problem, and (b) loss of performance due to multipath fading \cite{AKYILDIZ201140}. To overcome these problems, cooperation among CRs is desired \cite{5715894, AKYILDIZ201140}. The CSS method makes use of spatial diversity in CRs for an enhanced sensing accuracy. In CSS, the CRs can be simple sensors and the fusion centre (FC) can have a powerful computing capability. There is always a tradeoff between the CR overhead and the sensing accuracy. Classical fusion methods such as the OR rule, AND rule, K-out-of-N rule, counting rule, and linear quadratic combining rule have their associated advantages and disadvantages. These methods require setting a threshold value to make optimum decisions, and are not scalable. In CSS with ML framework, this threshold setting is not required.
 
Several ML algorithms, both supervised and unsupervised, have been proposed for CSS in existing literature \cite{6635250, 7564840, 9042894, 7751703, 7391780, 8466022}. Energy vector, where each element is the energy of the received signal measured by a CR has been used for classification \cite{8834099,6635250,5601105,8903028,9050784}. The training time of the ML models increases with increase in number of CRs. To overcome this problem, authors in \cite{7564840} propose using low dimensional probability vectors as features. It is shown that the $N$-dimensional energy vector $\mathbf{y}$ follows a multivariate Gaussian distribution. The mean vectors and covariance matrices under each hypotheses are written as
\begin{align}
    & \boldsymbol{\mu}_{y|H_{i}} = \left[\mu_{y_{1}|H_{i}},...,\mu_{y_{N}|H_{i}}\right]^{T}, ~~~i=0, 1,
\end{align}
\begin{align}
    & \boldsymbol{\Sigma}_{y|H_{i}}=\text{diag}\left \{\sigma^2_{y_{1}|H_{i}},...,\sigma^2_{y_{N}|H_{i}}\right \}, ~~~ i=0, 1,
\end{align}
where $\text{diag}(\cdot)$ denotes a diagonal matrix. The corresponding Gaussian distribution values of $\mathbf{y}$ are computed and fed as input to classifier, which reduces the training time. 

Finding an idle channel quickly with high probability can be improved, if the channel occupancy statistics by a PU is known. To this end, authors in \cite{9042894} use Q-learning to find the order in which channels are to be sensed. Further, to improve the sensing accuracy, discounted upper confidence bound (D-UCB) is used to select a SU partner for cooperation. The sensing accuracy is also known to increase if the data points of signal present and absent classes are well separated. The hidden PU problem is known to decrease the separability of these classes \cite{8662654}. To tackle this, \cite{8662654} proposes a strategy to place CRs to increase spectrum sensing coverage. The analytical expression for spectrum sensing coverage is derived based on Kullback-Leibler divergence between the distributions under each class.

Non-orthogonal multiple access (NOMA) technique is also known and is used to improve spectrum utilization. Authors in \cite{9102451} consider power-domain NOMA with CSS, in which the same resource block is used by two PUs with different transmit power levels. Unsupervised algorithms such as GMM, K-Means, and supervised algorithms such as K-nearest neighbor (KNN), directed acyclic graph support vector machine (DAG-SVM), and back propagation (BP) algorithms were studied, with energy vector as a feature. A total of $M$ CRs are considered, and each user measures energy locally over $L$ time slots. Thus a matrix $\mathbf{Y} \in L \times M$ is defined as a energy-based feature, which is considered for training and testing. Multiple PUs may operate in the environemnt around a CR, over the same frequency band. Such a setup is considered in \cite{7391780}, which considers a cellular network where frequency is reused by neighbouring cells, which are sufficient distance apart. In this setup, a multi-class classification is considered, by considering spectrum availability across space and time. Error-correcting output code based multi-class support vector machine classifier is used for classification.

In a large-scale heterogeneous CR network (CRN), a large number of static CRs are required for CSS, which is unfeasible. Additionally, in the case of mobile CRs, the availability of spectrum as perceived by CRs at different locations may be different, because of pathloss, shadowing, and fading. The work in \cite{8466022} proposes a Bayesian ML framework, where mobility of the CRs is exploited to collect CSS results and derive the global spectrum states. This model is used to capture the spatial-temporal correlation in the SS data. Later, Bayesian inference is used to group the sensing results without the prior knowledge of the number of states. Analysis and simulations for various CRN configurations are provided.

In a CSS scenario, the number of CRs may not be sufficient to sense all the PU channels for some applications. To solve this, a multiuser, multiband SS problem was formulated as a partially observable stochastic game in \cite{6507570}, which proposed a collaborative distributed multiagent RL-based SS to solve this problem. The CRs select a detection threshold such that a constraint on probability of miss is satisfied. The selection of the threshold is done according to the required diversity order. In each state, each SU chooses a frequency band to sense, and receives a reward based on the sensing results. The local decisions are fused by considering the neighbouring CRs for each frequency band. The reward each CR receives is based on the number of vacant channels. The aim of each CR is to increase the expected sum of future rewards given a constraint on the probability of miss. To optimize the above goal, an optimal action-value function is used. However, constructing lookup-tables of $q$ values across all possible state-action pairs is not practical. Therefore, \cite{6507570} employs a linear approximation on the action-value function, which reduces the computational complexity. Through simulation results, it is shown that this method provides an efficient way to find vacant spectra in a multiuser, multiband CRN. 

Deep learning (DL) on each CR to know about the status of the PU channel is presented in \cite{9050784}. A matrix containing locally measured energy in each frequency band by each CR is fed as input to the DL model. However, collecting this information from each SU increases network flow and usage of resources. Hence, reinforcement learning (RL) is used to select a few CRs to update their locally sensed energy in the input matrix in the present time slot. In RL, there is a trade-off between exploration and exploitation \cite{appl_rl}. The term exploitation refers to choosing an optimal action from the action space which provides maximum rewards, whereas exploration is choosing an action other than the present optimal action so as to explore new actions which may provide higher rewards. Exploitation-only causes convergence to a suboptimal solution, while frequent exploration decreases average rewards. Authors in \cite{8811461} address this tradeoff by considering the upper bound with Hoeffding-style (UCB-H) method, which improves the exploration efficiency. The CRN considered has $N_P$ PUs, $N_S$ CRs, and $M$ spectrum bands. Each CR can sense $K < M$ spectra in a given time slot, which is further divided into three mini slots namely, sensing, collaboration and access minislots. A separate collaboration channel exists for CRs to exchange decisions. Each CR performs SS and gathers information from other CRs and employs a learning algorithm for its sensing strategy. Two learning algorithms were considered, namely multi-agent Q-learning with UCB-H, and multi-agent deep q-network with UCB-H \cite{8811461}. 

\begin{figure}[t!]
    \centering
    \includegraphics[height=4cm, width=8cm]{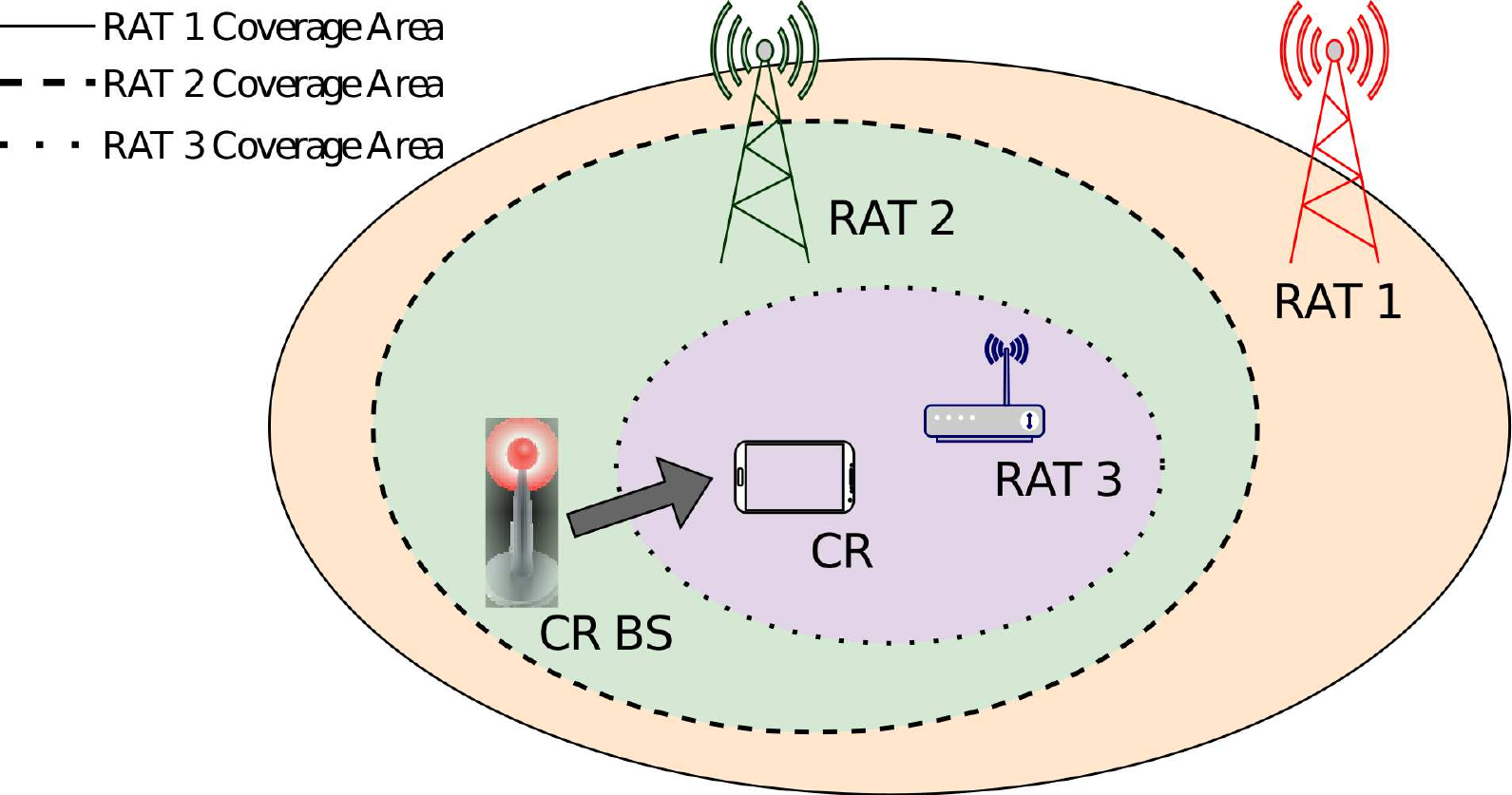}
    \caption{A CR surrounded by different RATs.}
    \label{fig:rat}
\end{figure}

\section{Dynamic Spectrum Allocation}\label{SecDSA}
After sensing the channels and detecting the spectrum holes, CRs are assigned a channel each for data transmission. Conventionally, fixed spectrum is assigned to users to perform data transfers. However, in a CRN, the CRs need to vacate the spectrum upon the arrival of the PU. Hence DSA is necessary to prevent interference with the PU. The main task in DSA is to allocate available channels to different CRs based on a chosen performance metric, with a constraint on interference to PUs and other CRs. 
\begin{figure*}[ht!]
\begin{subfigure}{.325\textwidth}
  \centering
  \includegraphics[height=5cm, width=7cm]{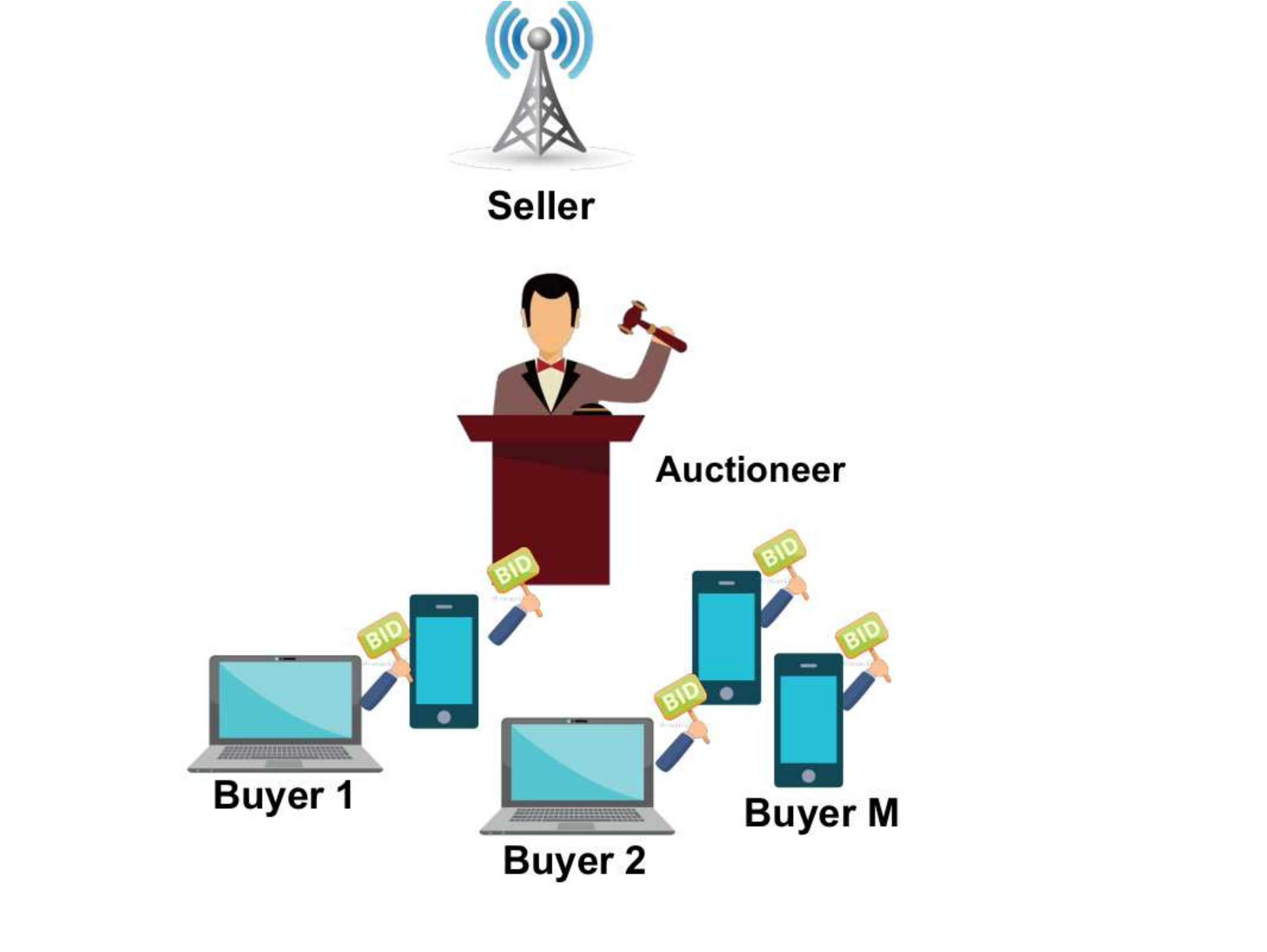}
  \caption{Forward auction with a single seller.}
  \label{fig:auc-first}
\end{subfigure}
\begin{subfigure}{.325\textwidth}
  \centering
  \includegraphics[height=5cm, width=7cm]{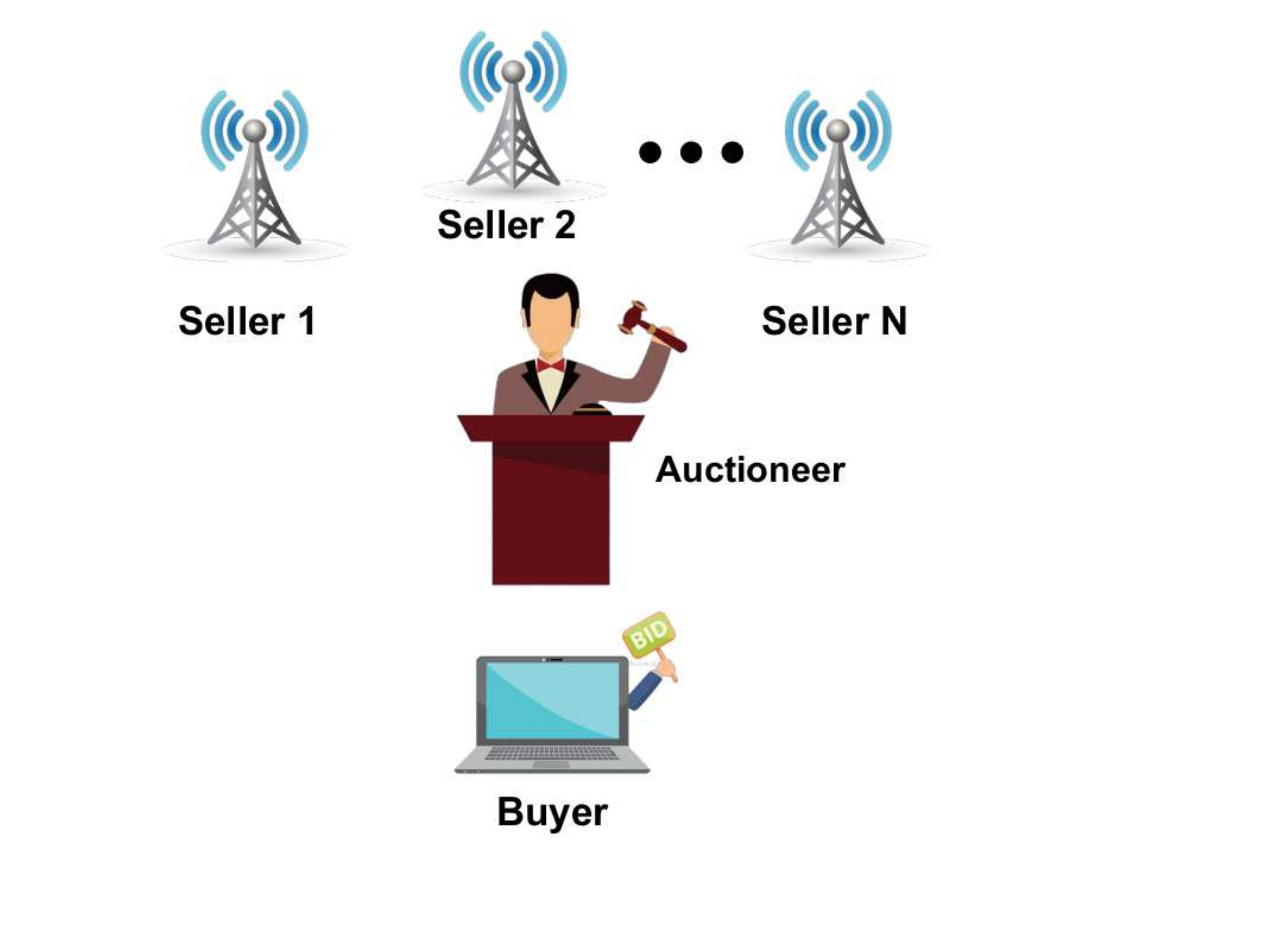}
  \caption{Auction with a single buyer.}
  \label{fig:auc-second}
\end{subfigure}
\begin{subfigure}{.325\textwidth}
  \centering
  \includegraphics[height=5cm, width=7cm]{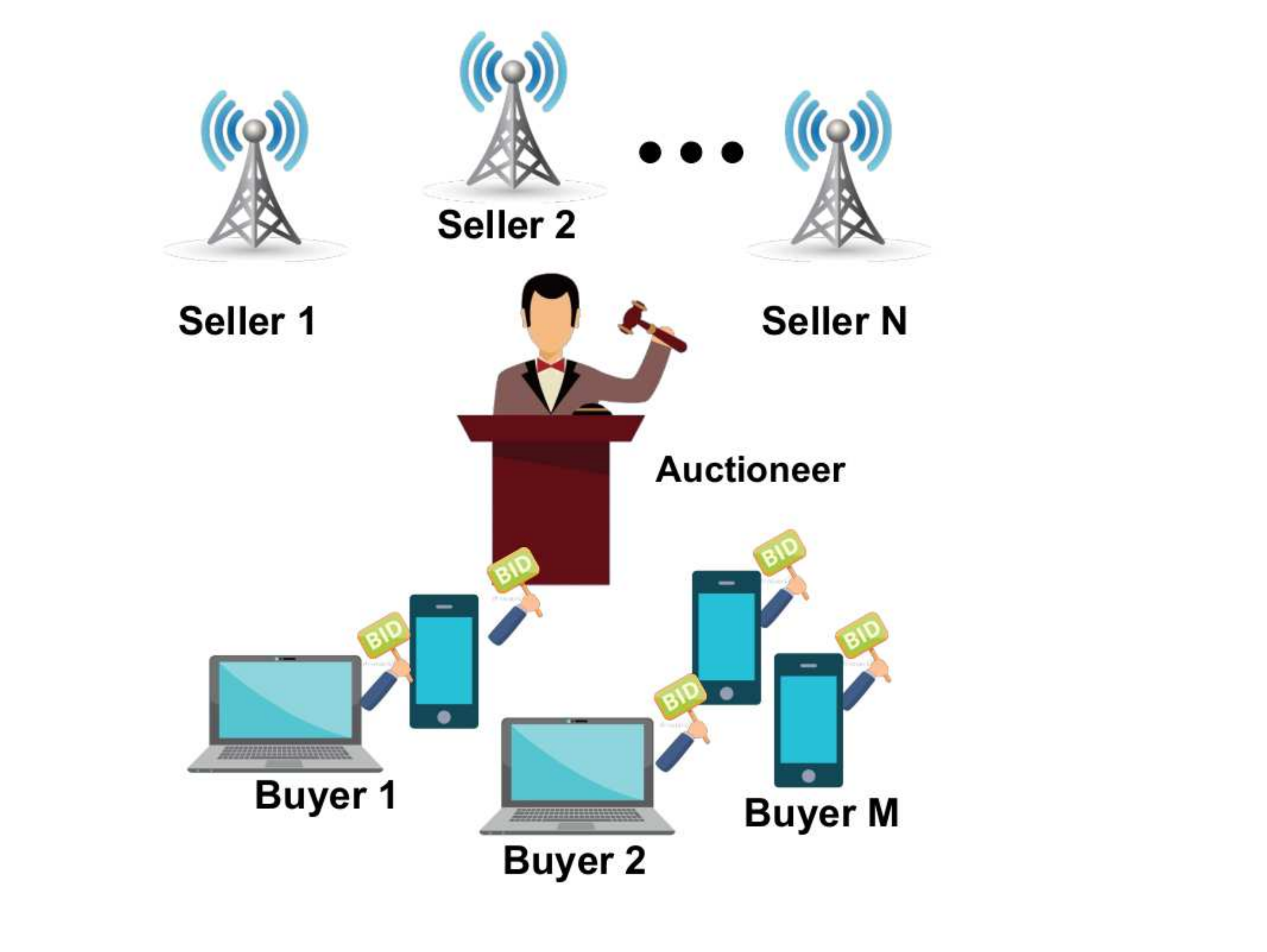}
  \caption{Auction with multiple buyers and sellers.}
  \label{fig:auc-third}
\end{subfigure}
\caption{Various types of auction models: (i) Single auction as shown in (a) and (b), and (ii) Double action as shown in (c).}
\label{fig:auction_models}
\hrulefill
\end{figure*}

In \cite{6276205}, DSA for a decentralized, ad-hoc CRN is considered. The CRs continuously learn their environment and share the information with neighboring CRs over a common channel. Each CR learns about the pattern of channel acquisition by the PU and other CRs, which is carried out by Tsigankov-Koulakov self-organizing maps, a special case of ANN. The throughput of CRs can be significantly improved if patterns in the PU traffic is exploited by a learning algorithm. In \cite{8270374}, CRs use a two-stage reinforcement learning to estimate the channel to be sensed, and its availability time. In the first stage, the modified optimistic Thompson sampling algorithm is used to return a list of available channels in order. In the second stage, a Bayesian method is employed to determine the length of channel availability. Authors in \cite{8454441} define a monopolistic market environment with a single PU (monopoly), $N$ CRs and $C$ subchannels. Each CR bids for channel access from a PU, and the PU gives channel access to the highest bidder. Probabilistic reinforcement learning is employed in both PU and CRs to learn the optimum strategy.

Authors in \cite{8720154} consider QoS requirements of CRs, and an optimization step is proposed before SS. Two strategies are proposed, namely wideband predictive sensing (WBPS) and Q-learning enabled WBPS (QWBPS). Prediction of future traffic loads of various radio access technologies (RAT) such as GSM, LTE, IEEE 802.11n, is performed at the CR base station (BS). An example of a SU surrounded by different RATs is shown in Fig.~\ref{fig:rat}. In WBPS, the BS chooses a RAT with less likely traffic load, based on the request by a CR node. Then, CR performs narrow band SS for the activity of that RAT. In QWBPS, the BS performs Q-learning by taking input as QoS requirements, demands of the CR node and predicted traffic loads to find the best RAT to sense. In WBPS, only the latency requirement of the CR node is considered, while in QWBPS, coverage and bandwidth requirements are also considered. 

\subsection{Spectrum Auction and Leasing}\label{SecAuction}
Spectrum auction is an effective method to exploit the under-utilized spectrum for data transmission. When not in use, the licensed spectra is auctioned to unlicensed users. In return, licensed users relay their transmission through the unlicensed users to save power. An extensive survey on the spectrum auction procedures for wireless systems is presented in \cite{6365159}. In-depth analysis of various auction models like forward, single-sided, double-sided, open-cry, and sealed cry is presented. Figure~\ref{fig:auction_models} shows the different type of such auction models. Further, a survey of single-hop and multi-hop systems are presented as examples for resource allocation in CRNs.

In DSA, only CRs are assumed to be the players. Each player can place a bid by specifying their will to spend certain amount of power to relay the PU data. In \cite{5928484}, an asymmetric cooperative communication-based spectrum leasing for both centralized and distributed architectures is presented. In the former, a decision centre assigns a bid for each PU channel based on the previous assignments through reinforcement learning. In the latter, an auction game-based protocol is presented, where a CR places a bid on every channel of the PU separately. The PUs on each channel selects a CR based on the bid that helps it to maximize the total power savings. The PU receivers can learn and adapt through reinforcement learning and increase the minimum bid to minimize their power spent for data transmission. 

Repeated auction has been adopted to maximize the trade-off between the gain and cost in accessing channels \cite{5928484,5692900}. CRs are selected based on their past auction history and valuations. Spectrum access in CR can be formulated as a dynamic game because of the distributed nature of the users. In \cite{5692900}, a Bayesian non-parametric belief update scheme is proposed to solve such a game. This scheme achieves a good trade-off between fairness and efficiency. A two-stage spectrum sharing method is presented in \cite{6924778}. They consider a scenario where each PU has a different amount of spectrum to sublet. The spectrum auction is performed in the first stage. Based on the winner, each PU determines a limit for the spectrum recall for the next round. In the next stage, using the Stackelberg pricing game, spectrum recall on both PUs and CRs is optimized. In the Stackelberg game, PUs and CRs act as leaders and followers respectively. Any strategy adopted by the PUs affects the actions of the CRs. Therefore PUs adopt an optimal strategy by studying the actions taken by the CRs. Spectrum is assigned to the winning CR. However, CRs are informed of the spectrum recall. Therefore, CRs start competing among themselves to reduce the total sum of recalled spectrum. The outcome of this process is reduction in payments. Both theoretical and simulation results show that using this two-stage resource allocation scheme in spectrum sharing improves the utilities.

The spectrum auctions among networks are proposed in \cite{8438331, 8462765}. In \cite{8438331}, the PUs are permitted to lease their unused spectrum to the CRs in multiple time slots. They propose a two-layer heterogeneous network in which a macro base station covers all the licensed and unlicensed users. CRs are covered by small cell base stations. A deep feed-forward network is used in the auction process. By adopting deep neural networks, the performance of the auction process was significantly improved when compared to the traditional auction methods. Social welfare was greatly improved by increasing the number of SUs, thereby raising the revenue of the PUs. Simulations were performed to prove this result. Spectrum auction framework for hybrid access in OFDMA-based cognitive macro-femto cell networks is proposed in \cite{8462765}. Similar to \cite{8438331, 8462765}, the work offloads some macro users to Femto access points to save energy. An incentive mechanism was proposed in which the spectrum was used as a reward for relaying PU data.

During the auction process, few users can use the spectrum excessively by lying about their real need for spectrum resources \cite{8434198}. As a result, the overall performance of the network and the service quality of other users is affected. Additionally, many CRs can resell the spectrum resources to other CRs. This collusion between the CRs affects the entire auction process. Authors in \cite{8434198} focus on suppressing sublease collusion, leasing the assigned channel to other CRs, which do not interfere to gain more revenue. By using the KNN algorithm to classify the users based on their geographical location and interference radius, buyers are categorized into virtual buyers and real buyers. Virtual buyers are users with no mutual interference, this will prevent the actual users to sublet the channels to them. 

\subsection{Spectrum Prediction}\label{SecSP}
Spectrum prediction is an important topic of research in the field of CR, which aims to improve the time and energy efficiencies of the SS process. SS with a large number of PU channels would require large sensing time and considerable sensing energy. With spectrum prediction, CRs minimize the number of channels sensing channels by choosing those which have a high chance to be vacant in the upcoming time slot. The prediction algorithms have become significantly better with the advent of ML research. This section aims to capture some of the key works in this aspect. In \cite{7746632}, four supervised ML techniques, recurrent neural networks, multilayer perceptron and SVM (Gaussian and linear kernel) were compared for DSA. Poisson, interrupted Poisson and self-similar traffics were used for analyzing the PU environment. Among the learning models considered, SVM with linear kernel was found to yield a consistently best performance across all the traffic patterns. The LSTM networks are a popular choice due to its feedback nature and ability to remember past inputs and results. The LSTM networks are designed to tackle the problem of long-term dependency. An LSTM network was used in \cite{8322623}, whose output was classified into two classes. With the same number of hidden layers and neurons, LSTM shows more encouraging results compared to the back-propagation network. LSTM networks are also used to utilize the spectral-temporal correlation among historical spectrum availability data to make predictions, such as in \cite{8001877}. Predictions on real-world occupancy of spectrum are made through analysis of time-frequency correlations using a LSTM \cite{8991968}, wherein the look back length of the LSTM network is established by analyzing auto-correlation function of spectrum occupancy. A soft cooperative fusion model based on LSTM is described in \cite{8539570} to capture both the spatial and temporal dependencies found in spectrum detection data. A long term prediction model was developed using convolution LSTM network with sequence-to-sequence architecture and was shown to take leverage of long sequences of spectrum data\cite{8902956}.

A deep learning-based approach was developed in \cite{8456449} by simultaneously predicting multi-slot ahead states of multiple spectrum points within a period of time. Supervised learning is initially employed to construct samples on short and long term high frequency data. Further, advanced residual network modules extract  characteristics in these data with diverse time scales, which are classified using a deep temporal-spectral residual network. A fusion of statistical and predictive modelling of spectrum occupancy was proposed in \cite{8600244} for enhanced dynamic spectrum access. An investigation was conducted to characterize the spectrum occupancy of the PU with the help of generalized Gaussian mixture models(GMMs). Further, the working of the proposed GMM was validated through learning based prediction via recurrent neural networks. Neural networks (NN) offer both flexibilities in architecture and consistency in results. Hence, a number of variants of NN are proposed and are shown to have reliable performance. A multilayer perceptron (MLP) is implemented in \cite{5502348}. The model does not require a prior knowledge of the traffic distributions of the PU and hence is particularly useful in real-life scenarios. In \cite{7393278}, sigma-if neural network is used which has a better efficiency and better performance than the MLP model. Another architecture that does not require any a priori knowledge is the neural network developed on the basis of a genetic algorithm model. Simulation results in \cite{7129617} indicate that such a method can predict spectrum availability accurately and further improve the prediction accuracy compared to other neural network-based prediction techniques. A further improvement to the conventional back propagation (BP) based NN is seen in \cite{7378129}. For input, channel states are used instead of channel power values. An optimization of the back propagation method is performed by using momentum algorithm and genetic algorithm in the prediction scheme. Further, a threshold interval is applied to determine predicted channel states. This is shown to have superior performance when compared to the conventional BP based NN. Time delay neural networks are designed to predict time-series data. In \cite{8717824}, authors use TDNN to represent the spectrum occupancy in land mobile radio bands (LMR). TDNN yield improvements over seasonal auto-regressive integrated moving average models in predicting short time horizons. TDNN can also be a good alternative for predicting spectrum vacancy in frequency bands that have similar characteristics of LMR channels.

A five-layer dynamic fuzzy neural network is designed in \cite{7852855}. The use of fuzzy theory helps in enhancing the error identification ability of the system and its robustness in dealing with certainty information and uncertainty information simultaneously. This was made possible by the parallel working of parameter adjustment and structure identification. A K-means clustering algorithm integrated with radial basis function (K-RBF) is proposed in \cite{7222923} and is shown to be an improvement over RBF. To evaluate the performance of machine-learning based spectrum prediction algorithms, \cite{6061664} devices two performance evaluation parameters based on packet loss rate and throughput. Next, we discuss some challenges encountered while employing ML methods in CRNs.

\section{Challenges} \label{SecChall}
The main challenges associated with ML algorithms in CRNs include: (a) complexity of the real-time model, and (b) the convergence of these algorithms within a limited time \cite{Nguyen2012}. Most of the ML algorithms learn from a pre-recorded dataset. They find patterns, develop understanding and evaluate their prediction confidence from the training data. Therefore the training time depends on the size of the dataset and the veracity of the dataset \cite{Abbas2015}. Another challenge faced is to ensure that the security of the device and the network is not comprised \cite{7084371}. As the number of devices grows, there is an increase in the device-to-device interaction in a distributed architecture. Therefore, PU emulation can send bogus SS results, thereby degrading the throughput of the network. In a centralized architecture, malicious users can overload the BS with erroneous results. The deployed algorithms have to provide a work around to mitigate this. Further, real-world test bed for testing these algorithms has not yet been systematically addressed \cite{7217798}. Hence, CR algorithms trained a dataset may not be robust to work effectively in a different environment. Robustness and scalability are two key problems which are encountered while implementing ML methods for CRNs.

\section{Conclusion}\label{conc}
In this paper, we discussed ML algorithms for CRNs for non-cooperative and cooperative SS, spectrum auction and prediction in DSA applications. At a high level, we discussed the set of major challenges in CRNs and mentioned the ML techniques to alleviate them, and also briefly discussed the challenges in implementation of these ML algorithms. Given the pros and cons of the discussed ML techniques, they offer a great value addition to performance analysis and practical implementation of CRNs in the near future.

\balance
\bibliographystyle{IEEEtran}
\bibliography{IEEEabrv,MLCRRefs}

\end{document}